# Impact of E-Banking on Traditional Banking Services

Shilpan Vyas

School of Computer Science and Information Technology,

Singhania University, Pacheri Bari, Jhunjhunu – 333515

Rajasthan, India.

*Abstract* - **Internet banking is changing the banking industry, having the major effects on banking relationships. Banking is now no longer confined to the branches were one has to approach the branch in person, to withdraw cash or deposit a cheque or request a statement of accounts. In true Internet banking, any inquiry or transaction is processed online without any reference to the branch (anywhere banking) at any time. Providing Internet banking is increasingly becoming a "need to have" than a "nice to have" service. The net banking, thus, now is more of a norm rather than an exception in many developed countries due to the fact that it is the cheapest way of providing banking services.**

**This research paper will introduce you to e-banking, giving the meaning, functions, types, advantages and limitations of e-banking. It will also show the impact of e-banking on traditional services and finally the result documentation.**

## I. INTRODUCTION

Internet banking (or E-banking) means any user with a personal computer and a browser can get connected to his bank's website to perform any of the virtual banking functions. In internet banking system the bank has a centralized database that is web-enabled. All the services that the bank has permitted on the internet are displayed in menu. Once the branch offices of bank are interconnected through terrestrial or satellite links, there would be no physical identity for any branch. It would be a borderless entity permitting anytime, anywhere and anyhow banking.

The network which connects the various locations and gives connectivity to the central office within the organization is called intranet. These networks are limited to organizations for which they are set up. SWIFT is a live example of intranet application.

E-banking provides enormous benefits to consumers in terms of ease and cost of transactions, either through Internet, telephone or other electronic delivery. Electronic finance (E-finance) has become one of the most essential technological changes in the financial industry. E-finance as the provision of financial services and markets using electronic communication and computation. In practice, e-finance includes e-payment, e-trading, and e-banking.

## II. MEANING OF E-BANKING

E-bank is the electronic bank that provides the financial service for the individual client by means of Internet.

## III. FUNCTIONS OF E-BANKING

At present, the personal e-bank system provides the following services: -

### A. INQUIRY ABOUT THE INFORMATION OF ACCOUNT

The client inquires about the details of his own account information such as the card's / account's balance and the detailed historical records of the account and downloads the report list.

### B. CARD ACCOUNTS' TRANSFER

The client can achieve the fund to another person's Credit Card in the same city.

### C. BANK-SECURITIES ACCOUNTS TRANSFER

The client can achieve the fund transfer between his own bank savings accounts of his own Credit Card account and his own capital account in the securities company. Moreover, the client can inquire about the present balance at real time.

### D. THE TRANSACTION OF FOREIGN EXCHANGE

The client can trade the foreign exchange, cancel orders and inquire about the information of the transaction of foreign exchange according to the exchange rate given by our bank on net.

### E. THE B2C DISBURSEMENT ON NET

The client can do the real-time transfer and get the feedback information about payment from our bank when the client does shopping in the appointed web-site.

### F. CLIENT SERVICE

The client can modify the login password, information of the Credit Card and the client information in e-bank on net.

### G. ACCOUNT MANAGEMENT

The client can modify his own limits of right and state of the registered account in the personal e-bank, such as modifying his own login password, freezing or deleting some cards and so on.

### H. REPORTING THE LOSS IF THE ACCOUNT

The client can report the loss in the local area (not nationwide) when the client's Credit Card or passbook is missing or stolen.

## IV. TYPES OF E-BANKING

A. Deposits, withdrawals, inter-account transfer and payment of linked accounts at an ATM;

B. Buying and paying for goods and services using debit cards or smart cards without having to carry cash or a cheque book;

C. Using a telephone to perform direct banking- make a balance enquiry, inter-account transfers and pay linked accounts;

D. Using a computer to perform direct banking- make a balance enquiry, inter-account transfers and pay linked

## V. ADVANTAGES OF E-BANKING

A. *Account Information:* Real time balance information and summary of day's transaction.

B. *Fund Transfer:* Manage your Supply-Chain network, effectively by using our online hand transfer mechanism. We can effect fund transfer on a real time basis across the bank locations.

C. *Request:* Make a banking request online.

D. *Downloading of account statements as an excel file or text file.*

E. *Customers can also submit the following requests online: Registration for account statements by e-mail daily / weekly / fortnightly / monthly basis.*

- Stop payment or cheques
- Cheque book replenishment
- Demand Draft / Pay-order
- Opening of fixed deposit account
- Opening of Letter of credit

F. *Customers can Integrate the System with his own ERP*

G. *Bill Payment through Electronic Banking*

H. *The Electronic Shopping Mall*

I. *Effecting Personal Investments through Electronic Banking*

J. *Investing in Mutual funds*

K. *Initial Public Offers Online*

## VI. LIMITATION OF E-BANKING

A. *Safety situations around ATMs.*

B. *Abuse of bank cards by fraudsters at ATMs.*

C. *Danger of giving your card number when buying on-line.*

## VII. IMPACT OF E-BANKING ON TRADITIONAL SERVICES

E-banking transactions are much cheaper than branch or even phone transactions. This could turn yesterday's competitive advantage - a large branch network - into a comparative disadvantage, allowing e-banks to undercut bricks-and-mortar banks. This is commonly known as the "beached dinosaur" theory.

E-banks are easy to set up, so lots of new entrants will arrive. 'Old-world' systems, cultures and structures will not encumber these new entrants. Instead, they will be adaptable and responsive. E-banking gives consumers much more choice. Consumers will be less inclined to remain loyal.

Portal providers are likely to attract the most significant share of banking profits. Indeed banks could become glorified marriage brokers. They would simply bring two parties together e.g. buyer and seller, payer and payee.

The products will be provided by monolines, experts in their field. Traditional banks may simply be left with payment and settlement business even this could be cast into doubt.

Traditional banks will find it difficult to evolve. Not only will they be unable to make acquisitions for cash as opposed to being able to offer shares, they will be unable to obtain additional capital from the stock market. This is in contrast to the situation for Internet firms for whom it seems relatively easy to attract investment.

E-banking is just banking offered via a new delivery channel. It simply gives consumers another service (just as ATMs did).

Experience in Scandinavia (arguably the most advanced e-banking area in the world) appears to confirm that the future is 'clicks and mortar' banking. Customers want full service banking via a number of delivery channels. The future is therefore 'Martini Banking' (any time, any place, anywhere, anyhow).

Traditional banks are starting to fight back.

The start-up costs of an e-bank are high. Establishing a trusted brand is very costly as it requires significant advertising expenditure in addition to the purchase of expensive technology (as security and privacy are key to gaining customer approval).

E-banks have already found that retail banking only becomes profitable once a large critical mass is achieved. Consequently many e-banks are limiting themselves to providing a tailored service to the better off.

E-Banking transaction needs some interface to communicate with banking customer. All the electronic transaction performs through some interfaces.

The electronic devices which perform interact with customers and communicate with other banking system is called electronic banking delivery channels.

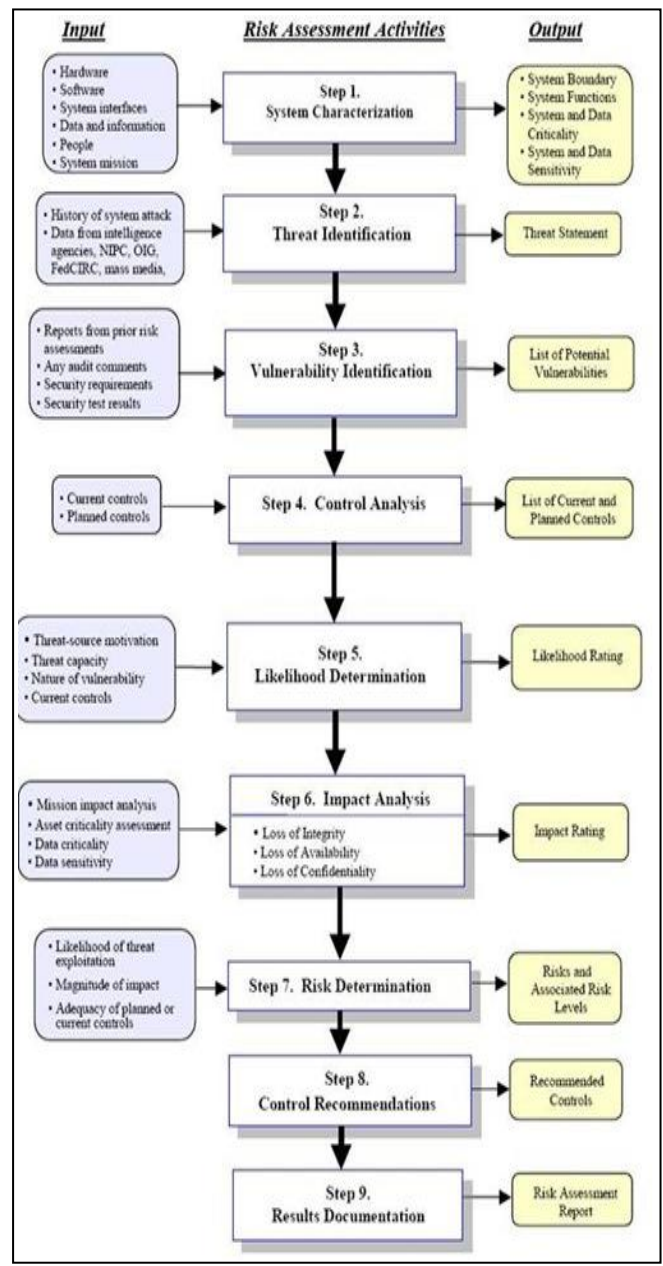

## VIII. RESULTS DOCUMENTATION

Once the risk assessment has been completed (threat-sources and vulnerabilities identified, risks assessed, and recommended controls provided), the results should be documented in an official report or briefing.

## IX. Conclusions

E-banking is a borderless entity permitting anytime, anywhere and anyhow banking. This facilitates us with all the functions and many advantages as compared to traditional banking services. During this step of the process, controls that could mitigate or eliminate the identified risks, as appropriate to the organization's operations, are provided. The goal of the recommended controls is to reduce the level of risk to the IT system and its data to an acceptable level.